# Cavity-enhanced Raman emission from a single color center in a solid

Shuo Sun[1‡*], Jingyuan Linda Zhang[1‡], Kevin A. Fischer[1‡], Michael J. Burek[2], Constantin Dory[1], Konstantinos G. Lagoudakis[1], Yan-Kai Tzeng[3], Marina Radulaski[1], Yousif Kelaita[1], Amir Safavi-Naeini[1], Zhi-Xun Shen[3,4,5], Nicholas A. Melosh[4,5], Steven Chu[3,6], Marko Lončar[2], and Jelena Vučković[1]

[1]E. L. Ginzton Laboratory, Stanford University, Stanford, California 94305, USA
[2]School of Engineering and Applied Sciences, Harvard University, Cambridge, Massachusetts 02138, USA
[3]Department of Physics, Stanford University, Stanford, California 94305, USA
[4]Geballe Laboratory for Advanced Materials, Stanford University, Stanford, California 94305, United States
[5]Stanford Institute for Materials and Energy Sciences, SLAC National Accelerator Laboratory, Menlo Park, California 94025, USA
[6]Department of Molecular and Cellular Physiology, Stanford University, Stanford, California 94305, USA
[‡]These authors contributed equally
[*]Email: shuo@stanford.edu

**Abstract**

We demonstrate cavity-enhanced Raman emission from a single atomic defect in a solid. Our platform is a single silicon-vacancy center in diamond coupled with a monolithic diamond photonic crystal cavity. The cavity enables an unprecedented frequency tuning range of the Raman emission (100 GHz) that significantly exceeds the spectral inhomogeneity of silicon-vacancy centers in diamond nanostructures. We also show that the cavity selectively suppresses the phonon-induced spontaneous emission that degrades the efficiency of Raman photon generation. Our results pave the way towards photon-mediated many-body interactions between solid-state quantum emitters in a nanophotonic platform.

Integration of solid-state quantum emitters with nanophotonic structures offers a scalable quantum photonics platform[1] that is essential for photonic quantum simulation[2], quantum metrology[3], quantum repeaters[4], and quantum networks[5,6]. However, despite significant progress in coupling single solid-state qubits with photons[7-10] and entangling two qubits[11-14], a scalable quantum photonic circuit consisting of many quantum emitters remains an outstanding challenge. One major obstacle towards this goal is the spectral inhomogeneity of solid-state quantum emitters[15], which limits their prospects in realizing many-body interactions through exchange of photons[5]. The ability to tune the emission frequency of a solid-state quantum emitter across the full range of inhomogeneous broadening remains a key missing ingredient in developing scalable quantum photonic circuits.

Color centers in solids have recently shown great promise for applications in scalable quantum photonic circuits, largely owing to their narrow spectral inhomogeneity. One of the candidates that has attracted significant interests in recent years is the negatively charged silicon-vacancy ($SiV^-$) center in diamond. $SiV^-$ centers possess narrow inhomogeneous broadening on the order of 1 GHz in high quality diamond[16,17]. They also exhibit properties that make them promising as optically accessible quantum memories, including high spectral stability[16], large zero-phonon-line emission (>70%)[18], gigahertz coupling strength with nano-cavities[13,19], as well as milliseconds spin coherence time[20]. Recent experiments have demonstrated photon-mediated entanglement between two $SiV^-$ centers in a bare waveguide[13], where Raman emissions with a tuning range of 10 GHz were employed to compensate the spectral inhomogeneity of $SiV^-$ centers. However, there are two main limitations in using this approach towards realizing photon-mediated many-body interactions. First, once embedded in nanostructures, $SiV^-$ centers typically display a much larger spectral inhomogeneity (>20 GHz) than bulk due to the difficulties in engineering a homogeneous strain distribution[21]. Second, the observed Raman emission is accompanied with a strong spontaneous emission from the same branch of the Λ-system[13], which fundamentally limits the efficiency of Raman photon generation and the fidelity of many-body interactions. To address both challenges requires selective enhancement of the Raman emission while suppressing the undesired spontaneous emission.

In this Letter, we demonstrate cavity-enhanced Raman emission from a single color center. Cavity-enhanced Raman emission has been first demonstrated with single trapped atoms[22-24], where tuning of the emission frequency by ~100 linewidths has been achieved[25], much larger than trap-induced linewidth broadening. In solid-state platforms, optical cavities have been utilized to enhance Raman emission from a single quantum dot, which enables generation of single-photons with large tuning bandwidth[26] and variable pulse shape[27,28]. However, the cavity-enhanced tuning range remains two orders of magnitude smaller compared with the spectral inhomogeneity of quantum dots[29]. Here, we show that an optical cavity enables a frequency tuning range of 100 GHz for Raman emission from a single SiV$^-$ center in diamond, which is an order of magnitude larger than previously achieved with color centers and far exceeds the typical spectral inhomogeneity of SiV$^-$ centers in nanostructures. In addition, we provide a quantitative model to explain the undesired spontaneous emission by accounting for electron-phonon interactions, and show that the cavity can selectively suppress the spontaneous emission and only enhance the Raman photon generation. Our results represent an important step towards the implementation of scalable quantum circuits and quantum networks that involve multiple solid-state quantum emitters in an integrated nanophotonic platform.

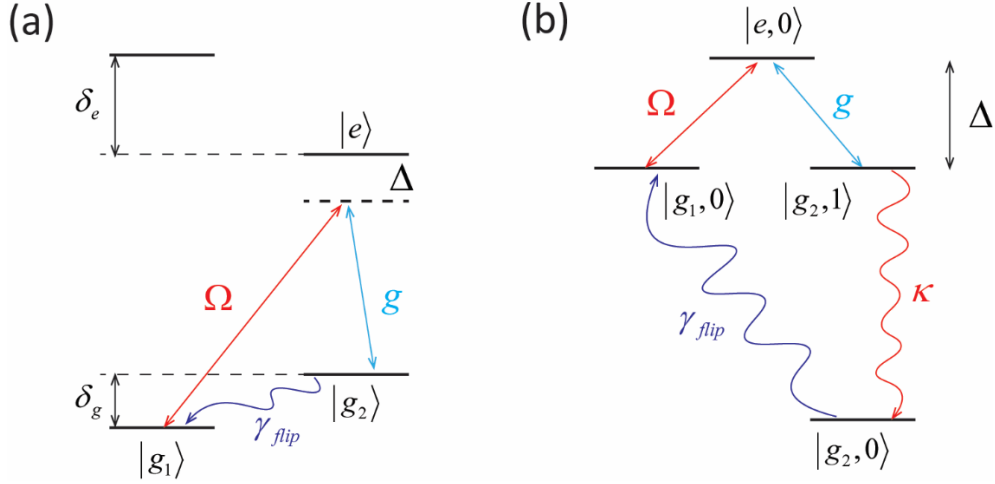

FIG. 1. (Color online) (a) Energy level structure of a SiV⁻ center. (b) Relevant energy level structure of the emitter-cavity system for cavity-enhanced Raman emission.

Figure 1(a) shows the energy level structure of a single SiV⁻ center[30]. In the absence of a magnetic field, the SiV⁻ center contains two ground states separated by $\delta_g$, and two excited states separated by $\delta_e$. The values of $\delta_g$ and $\delta_e$ are typically $\delta_g/2\pi = 50$ GHz and $\delta_e/2\pi = 260$ GHz respectively[30], but they increase significantly in the presence of strain[31,32]. In Fig. 2(c) we will show that for the specific emitter we measured, the ground state splitting is $\delta_g/2\pi = 544$ GHz. We utilize the Λ-system formed by the lower excited state (labeled as $|e\rangle$) and the two ground states (labeled as $|g_1\rangle$ and $|g_2\rangle$) to generate tunable Raman emission. We optically drive transition $|g_1\rangle \leftrightarrow |e\rangle$ using a continuous-wave laser with a Rabi frequency given by $\Omega$, and couple transition $|g_2\rangle \leftrightarrow |e\rangle$ to a cavity with a coupling strength given by $g$ (vacuum Rabi frequency of $2g$). We set the detuning between the driving laser and transition $|g_1\rangle \leftrightarrow |e\rangle$ to be identical to the detuning between the cavity and transition $|g_2\rangle \leftrightarrow |e\rangle$ (both are given by $\Delta$) in order to achieve Raman resonance[22-24]. Note that unlike the scheme of stimulated Raman adiabatic passage[33] that requires two lasers to drive both branches of the Λ-system, here we only need a single laser to drive one branch since the cavity will stimulate the emission from the other one.

To understand how we generate cavity-enhanced Raman emission, we illustrate the level structure in the interaction picture as shown in Fig. 1(b). We denote each state in the form $|x,n\rangle$, where $x \in \{g_1, g_2, e\}$ is the state of the SiV$^-$ center, and $n \in \{0,1\}$ is the number of photons in the cavity. By truncating the infinite Jaynes-Cummings ladders, we implicitly assume that the system contains at most one excitation. This assumption is always valid in the absence of ground state relaxation[34]. When accounting for ground state relaxation, this assumption corresponds to the condition $\kappa \gg \gamma_{flip}$, where $\kappa$ is the cavity energy decay rate, and $\gamma_{flip}$ is the ground state relaxation rate from $|g_2\rangle$ to $|g_1\rangle$. We also assume that $\Omega, g \ll \Delta$, so that we can adiabatically eliminate the state $|e,0\rangle$, and treat the system as two-levels $|g_1,0\rangle$ and $|g_2,1\rangle$ driven by an effective Rabi frequency $\Omega_{eff} = \Omega g / \Delta$ [34]. Thus, if the system is initially in the state $|g_1,0\rangle$, it will coherently rotate to the state $|g_2,1\rangle$ with a Rabi frequency $\Omega_{eff}$, which then decays to the state $|g_2,0\rangle$ via emitting a photon through the cavity. The emission frequency is tunable with $\Delta$ because it does not involve any real excitation of the state $|e,0\rangle$. We utilize the phonon-mediated ground state relaxation to reinitialize the state from $|g_2,0\rangle$ back to $|g_1,0\rangle$ after the Raman emission. Note that the reverse relaxation process from $|g_1,0\rangle$ to $|g_2,0\rangle$ is negligible as has been demonstrated recently[31], because it requires absorption of a single phonon at the frequency $\delta_g/2\pi = 544$ GHz, which is much larger than the thermal energy $k_B T = 83$ GHz at the measurement temperature of 4 K.

The coupling between the emitter and the cavity enhances the rate of the Raman emission.

Here we define the Raman emission rate as the inverse of the average time it takes to emit a photon when the system is initially in the state $|g_1, 0\rangle$. In Supplementary Materials[34], we demonstrate that the cavity-enhanced Raman emission rate is given by $R_c = \frac{\Omega_{eff}^2}{\kappa} = \frac{g^2}{\kappa} \cdot \left(\frac{\Omega}{\Delta}\right)^2$, while the upper bound of the Raman emission rate without a cavity is given by $R_0 = \frac{(\Omega/2)^2}{\Delta^2 + (\Gamma/2)^2} \cdot \Gamma$, where $\Gamma$ is the spontaneous emission rate of transition $|e\rangle \to |g_2\rangle$. In the limit where $\Delta \gg \Gamma$, the Raman emission rate is enhanced by a factor $\frac{R_c}{R_0} = \frac{4g^2}{\kappa\Gamma}$, which is the Purcell factor of the coupled emitter-cavity system. For SiV$^-$ centers, the Purcell factor can be more than a factor of 10[19], corresponding to at least an order of magnitude enhancement of the Raman emission rate.

We couple a single SiV$^-$ center with a monolithic diamond nanobeam photonic crystal cavity[19]. Figure 2(a) shows a scanning electron microscope image of the fabricated cavity. The device fabrication starts with homoepitaxial growth of a thin layer of diamond on a single-crystal diamond substrate using microwave plasma chemical vapor deposition (MPCVD). We place a silicon wafer underneath the diamond substrate to generate silicon atoms in the growth chamber through hydrogen plasma etching, which then form SiV$^-$ centers due to plasma diffusion. We then fabricate nanobeam photonic crystal cavities using electron beam lithography followed by angled etching of the bulk diamond to create a suspended nanobeam[38].

We mount our sample in a closed-cycle cryostat and cool it down to 4 K. Supplemental Materials contain detailed descriptions of the measurement methods[34]. We first measure the bare cavity transmission spectrum using a supercontinuum source (Fig. 2(b)). By fitting the measured data (blue circles) to a Lorentzian function (red solid line), we obtain a cavity energy decay rate of $\kappa/2\pi = 53.7 \pm 0.4$ GHz (corresponding to a quality factor of 7600).

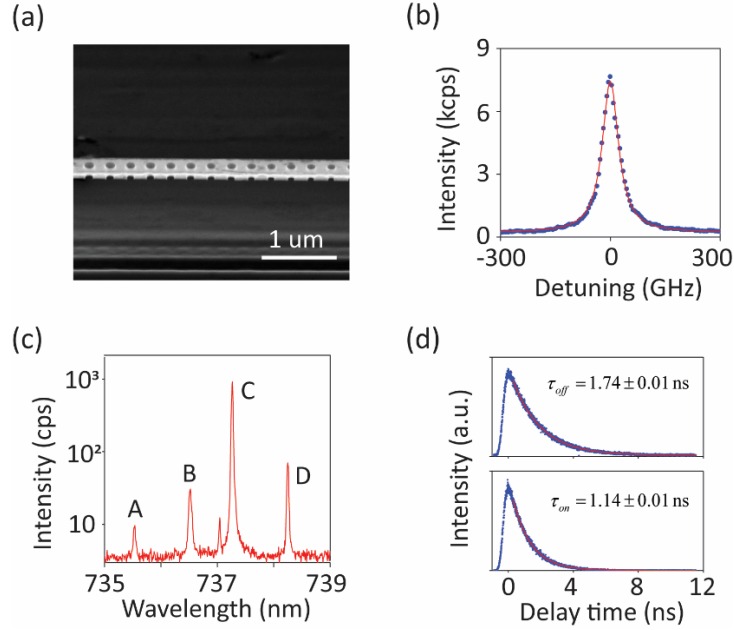

FIG. 2. (Color online) (a) Scanning electron microscope image of a fabricated nanobeam photonic crystal cavity in diamond. (b) Transmission spectrum of a bare cavity measured using a supercontinuum source. (c) Photoluminescence spectrum of the SiV$^-$ center we used in our experiment. (d) Lifetime measurement of the lower excited state of the SiV$^-$ center when the cavity is far detuned from the emitter (upper panel) and when the cavity is resonantly coupled with transition $|g_2\rangle \leftrightarrow |e\rangle$ (lower panel). In both panels (b) and (d), blue dots show the measured data, the red solid lines show the numerical fit.

Figure 2(c) shows the photoluminescence spectrum of the SiV$^-$ center embedded in the cavity. To eliminate the effect of the cavity on the emission spectrum, we red detune the cavity by more than 40 linewidths from all transitions of the SiV$^-$ center. We observe four distinct peaks in the photoluminescence spectrum, labeled as A – D in the figure, corresponding to the four optical transitions of a single SiV$^-$ center. The peaks C and D correspond to transitions $|g_1\rangle \leftrightarrow |e\rangle$ and $|g_2\rangle \leftrightarrow |e\rangle$ respectively[39]. From the frequency splitting between the emission peaks C and D, we

calculate that $\delta_g/2\pi = 544\,\text{GHz}$. This value is significantly larger than the value obtained in the bulk (50 GHz) using the same sample[34], suggesting large residual strain in the nanobeam photonic crystal. Second order correlation measurements verify that the emissions from both peaks C and D exhibit clear anti-bunching and are therefore originated from a single SiV$^-$ center[34]. We attribute the weak emission peak near transition C to a different emitter.

To characterize the coupling strength $g$ between the cavity and transition $|g_2\rangle \leftrightarrow |e\rangle$, we measure the lifetime of the excited state $|e\rangle$ both when the cavity is far detuned and resonant with the transition $|g_2\rangle \leftrightarrow |e\rangle$, as shown in the upper and lower panels of Fig. 2(d). By fitting the measured data (blue dots) to an exponential function (red solid line), we determine the lifetime of the excited state $|e\rangle$ to be $\tau_{off} = 1.74 \pm 0.01\,\text{ns}$ for the far detuned case, and $\tau_{on} = 1.14 \pm 0.01\,\text{ns}$ for the resonant case. We thus calculate the coupling strength to be $g/2\pi = 0.80 \pm 0.01\,\text{GHz}$ using the relation $\frac{1}{\tau_{on}} = 4g^2/\kappa + \frac{1}{\tau_{off}}$. We also estimate a lower-bound Purcell factor of 20[34].

We now demonstrate cavity-enhanced Raman emission. We excite the transition $|g_1\rangle \leftrightarrow |e\rangle$ using a continuous-wave laser with a variable detuning $\Delta$, and collect the emission from the cavity. To reject the direct reflection of the laser from the sample surface, we spatially separate the excitation and collection by irradiating the laser on a notch located at the end of the nanobeam, which is designed for coupling light from free-space to the waveguide[13,34]. We collect the far-field scattered signal from the cavity at the center of the nanobeam. We also use a double monochromator to further filter out the laser reflection and spectrally select the emission around transition $|g_2\rangle \leftrightarrow |e\rangle$ within a bandwidth of 120 GHz.

Figure 3(a) shows the measured emission spectrum as we vary the detuning $\Delta$. We observe two distinct peaks in the measured spectra, labeled as *R* and *S* respectively. The emission peak *R* continuously red shifts as we increase the detuning $\Delta$, corresponding to the cavity-enhanced Raman emission. The emission peak *S* remains centered around the natural frequency of $|g_2\rangle \leftrightarrow |e\rangle$, which is originated from incoherent excitation of the system into the state $|e\rangle$ followed by spontaneous emission via transition $|e\rangle \rightarrow |g_2\rangle$. We are able to achieve a tuning range of 99 GHz for the Raman emission, which is an order of magnitude larger than the best value achieved previously for a color center[13]. Note that the demonstrated tuning range is only limited by the bandwidth of our spectral filter (120 GHz), and does not constitute a fundamental limit.

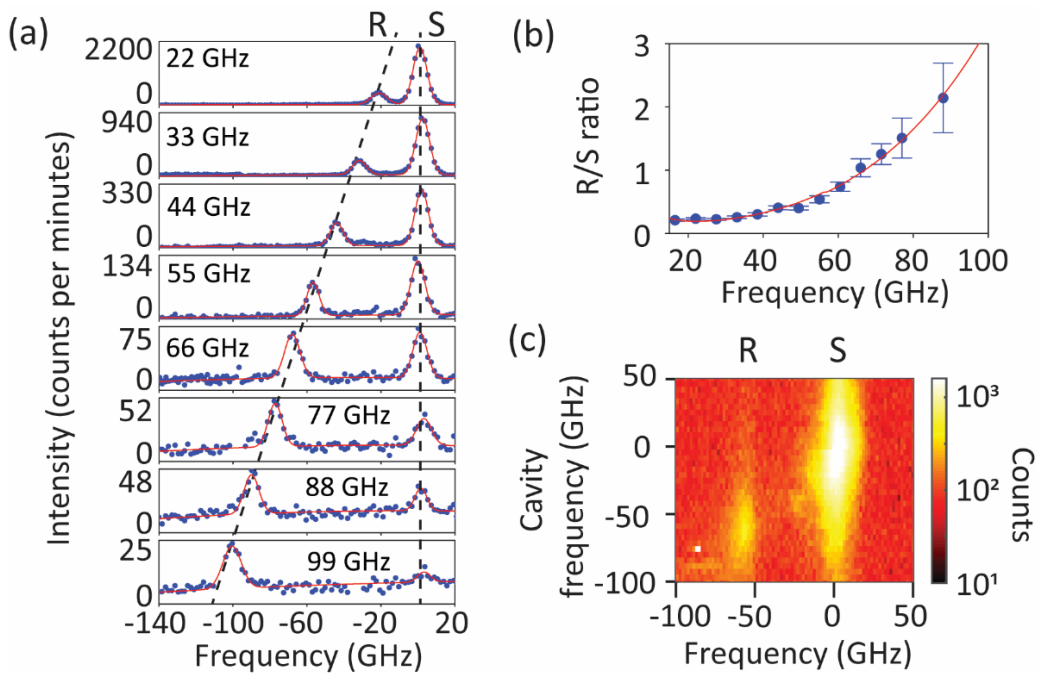

FIG. 3. (Color online) (a) Cavity emission spectra as we vary the excitation detuning $\Delta$. The blue dots show measured data, and the red solid lines show the numerical fits to a double Lorentzian function. The labels R and S represent the Raman and spontaneous emission peaks, respectively. (b) Ratio between the Raman and spontaneous emission intensity as we vary the excitation detuning $\Delta$. The blue circles show measured values, and the red solid line shows numerically calculated ratios. (c) Cavity emission spectra as we tune the cavity across both the spontaneous and Raman emission

peaks. In both panels (a) and (c), the frequency values are given in terms of detuning from transition $|g_2\rangle \leftrightarrow |e\rangle$.

Besides an unprecedented tuning bandwidth, the cavity also enables selective enhancement of Raman emission as we spectrally detune the Raman emission away from the emitter resonance. To quantitatively show this effect, we extract the ratio between the Raman and spontaneous emission intensity (referred as the R/S ratio) at each detuning, as shown in Fig. 3(b). The R/S ratio increases by a factor of 10 when we increase the detuning from 15 GHz to 88 GHz. The R/S ratio achieves even higher value at 99 GHz, but we cannot accurately calculate the ratio at this condition due to the vanishing spontaneous emission peak that is too close with the noise floor.

We now verify that the selective enhancement at large detuning originates from the cavity. We fix the excitation detuning at $\Delta/2\pi = 55$ GHz, and finely tune the cavity frequency across both the Raman and spontaneous emission peaks. If the improvement of R/S ratio at large detuning is not related with the cavity, we should observe no dependence of the R/S ratio as we sweep the cavity frequency. In contrast, as shown in Fig. 3(c), when the cavity is resonant at the Raman emission frequency (-55 GHz), we observe at least 10-fold enhancement of the Raman emission intensity compared with the case when the cavity is detuned 100 GHz away from the Raman emission. The cavity can also enhance the spontaneous emission, but at a different frequency (~0 GHz). These results confirm that the selective enhancement of the Raman emission is enabled by the cavity.

Finally, we investigate the origin of the strong spontaneous emission, especially at small detuning. In fact, previous studies have observed similar spontaneous emission[13], but the physical mechanism for this observation has not been explored thoroughly. We quantitatively explain the spontaneous emission by accounting for interactions between the $SiV^-$ center and a phonon reservoir. Specifically, we derive a microscopic model that quantifies how the state $|e\rangle$ is excited by absorbing both a photon from the driving field and a phonon from the reservoir, leading to the spontaneous emission.

We start with the Hamiltonian of the driven Λ-system shown in Fig. 1(b), given by

$$\hat{H}_{sys} = \Delta |e,0\rangle\langle e,0| + \left(\frac{\Omega}{2}|e,0\rangle\langle g_1,0| + g|e,0\rangle\langle g_2,1| + h.c.\right). \tag{1}$$

We model the phonons as a bath of harmonic oscillators, given by

$$\hat{H}_{bath} = \sum_{\mathbf{k}} \omega_{\mathbf{k}} \mathbf{b}_{\mathbf{k}}^{\dagger} \mathbf{b}_{\mathbf{k}}. \tag{2}$$

In Eq. (2), $\mathbf{k}$ is the wavevector of each phonon mode, $\omega_{\mathbf{k}}$ is the frequency of the phonon mode $\mathbf{k}$, and $\mathbf{b}_{\mathbf{k}}$ is the bosonic annihilation operator for the phonon mode $\mathbf{k}$. The interaction Hamiltonian between the SiV⁻ center and phonons could be written as

$$\hat{H}_{sys-bath} = \sum_{\mathbf{k}} \left(\mathbf{b}_{\mathbf{k}} + \mathbf{b}_{\mathbf{k}}^{\dagger}\right)\left(p_{\mathbf{k}}|g_1,0\rangle\langle g_1,0| + q_{\mathbf{k}}|g_2,1\rangle\langle g_2,1| + r_{\mathbf{k}}|e,0\rangle\langle e,0|\right), \tag{3}$$

where $p_{\mathbf{k}}$, $q_{\mathbf{k}}$, and $r_{\mathbf{k}}$ are the deformation coupling strength between the phonon mode $\mathbf{k}$ and the electronic states $|g_1\rangle$, $|g_2\rangle$, and $|e\rangle$ respectively. Note that here we do not include the phonon-induced ground state relaxation since this process only determines the number of excitation and emission cycles per second and does not affect the R/S ratio. We will add this term phenomenologically in the final master equation[34].

We now derive the electron-phonon interactions in the form of Lindblad operators following a similar formalism used for semiconductor quantum dots[40,41]. To derive the Lindblad operators, we first transform the interaction Hamiltonian $\hat{H}_{sys-bath}$ into the diagonal basis of $\hat{H}_{sys}$ (Eq. (1)), and then write it in the rotating reference frame with respect to $\hat{H}_{sys} + \hat{H}_{bath}$. The final master equation is given by $d\rho_{sys}/dt = -i\left[\hat{H}_{sys}, \rho_{sys}\right] + L_{phonon}(\rho_{sys})$, where $\rho_{sys}$ is the density matrix of the system, and $L_{phonon}(\rho_{sys})$ is the phonon dissipator, given by

$$L_{phonon}(\rho_{sys}) = \frac{g^2+(\Omega/2)^2}{\Delta^2}J_1(\Delta)\left[n_{th}(\Delta)D(|+\rangle\langle-|)+(1+n_{th}(\Delta))D(|-\rangle\langle+|)\right]$$
$$+\frac{g^2+(\Omega/2)^2}{\Delta^2}J_2(\Delta)\left[n_{th}(\Delta)D(|+\rangle\langle d|)+(1+n_{th}(\Delta))D(|d\rangle\langle+|)\right], \quad (4)$$

where $D(\hat{\mathbf{O}})\rho_{sys} = \hat{\mathbf{O}}\rho_{sys}\hat{\mathbf{O}}^\dagger - \frac{1}{2}\hat{\mathbf{O}}^\dagger\hat{\mathbf{O}}\rho_{sys} - \frac{1}{2}\rho_{sys}\hat{\mathbf{O}}^\dagger\hat{\mathbf{O}}$ is the general Lindblad superoperator for the collapse operator $\hat{\mathbf{O}}$. Note that here we only elaborate the phonon mediated dissipation for the convenience of discussion. The Supplementary Materials contain the complete master equation and detailed derivations[34]. In Eq. (4), the states $|+\rangle$, $|-\rangle$, and $|d\rangle$ are eigenstates of $\hat{\mathbf{H}}_{sys}$, given by

$$|+\rangle = \frac{\Omega}{2\Delta}|g_1,0\rangle + \frac{g}{\Delta}|g_2,1\rangle + |e,0\rangle, \quad (5)$$

$$|-\rangle = \frac{\Omega/2}{\sqrt{g^2+(\Omega/2)^2}}|g_1,0\rangle + \frac{g}{\sqrt{g^2+(\Omega/2)^2}}|g_2,1\rangle - \frac{\sqrt{g^2+(\Omega/2)^2}}{\Delta}|e,0\rangle. \quad (6)$$

$$|d\rangle = \frac{g}{\sqrt{g^2+(\Omega/2)^2}}|g_1,0\rangle - \frac{\Omega/2}{\sqrt{g^2+(\Omega/2)^2}}|g_2,1\rangle. \quad (7)$$

The parameters $J_1(\Delta)$ and $J_2(\Delta)$ are the spectral density of phonons that couple with the transition $|+\rangle \leftrightarrow |-\rangle$ and $|+\rangle \leftrightarrow |d\rangle$ respectively. The parameter $n_{th}(\Delta)$ is the number of phonons per mode, which follows the Bose-Einstein distribution given by $n_{th}(\Delta) = \left[\exp(\Delta/k_B T)-1\right]^{-1}$.

The phonon dissipator in the form of Eq. (4) has a clear physical intuition. It shows how the system can be populated incoherently into the dressed state $|+\rangle$ from the states $|-\rangle$ or $|d\rangle$ by absorption of a single phonon from the reservoir. Since $|+\rangle \approx |e\rangle$ in the limit $\Omega, g \ll \Delta$, the incoherent population transfer into the state $|+\rangle$ leads to spontaneous emission from the excited state. Eq. (4) also includes the reverse process where the state $|+\rangle$ decays to the states $|-\rangle$ or $|d\rangle$ by emitting a phonon, but this process has a minor effect since its rate is typically much slower than other decay mechanisms of the excited state $|+\rangle$.

We numerically solve the master equation of the system, and calculate the cavity emission spectrum using the quantum regression theorem[34]. We set all the parameters using experimentally measured values, except for the phonon spectral densities $J_1(\Delta)$ and $J_2(\Delta)$. The exact form of $J_1(\Delta)$ and $J_2(\Delta)$ depends on many parameters such as the strain susceptibility of each electronic state of the SiV$^-$ center, the local strain of each phonon mode, and the phonon frequency dispersion, which is difficult to derive from the first principles. Here, we qualitatively assume a phonon spectral density function of the form $J_{1,2}(\Delta) = \alpha_{1,2}\Delta^n$, where $\alpha_{1,2}$ is a trivial scalar, and $n$ represents a geometric scaling factor that is determined by the structure[31]. For example, for phonons in the bulk $n = 3$, but for surface phonons $n = 2$. The red solid line in Fig. 3(b) shows the calculated R/S ratio using our model. For the best fit, we obtain $n = 0.31 \pm 0.24$. This value is much smaller than the bulk value of 3, suggesting that the nanobeam strongly modifies the phonon spectral density.

In conclusion, we have demonstrated cavity-enhanced Raman emission from a single SiV$^-$ center. The cavity enables an unprecedented frequency tuning range of 99 GHz, which significantly exceeds the typical spectral inhomogeneity of SiV$^-$ centers in nanostructures. We also demonstrate that the cavity selectively enhances only the Raman emission, which is critical for achieving high-fidelity photon-mediated many-body interactions. In our current experiment, we employed two orbital ground states to form a $\Lambda$-system, which have short lifetimes[37] and thus limit our capability to generate single photons due to fast re-excitation. In order to obtain pure single photons from the Raman emission, we could utilize the spin sublevels of SiV$^-$ centers, which have lifetimes of milli-seconds at cryogenic temperature[42,43] and seconds at milli-Kelvin temperature[20]. The long coherence time of the electron spin may further enable quantum state transfer between single spins and photons through cavity stimulated adiabatic Raman passage[22]. Another important property for photon-mediated many-body interactions is the photon coupling efficiency. In our current device, the input and output photon coupling are achieved through free-space scattering from either the cavity or the notches at the end of the waveguide, which has a limited efficiency on the order of 1%[13]. Such coupling efficiencies can be significantly improved by using an adiabatic tapered fiber to directly couple with the nanobeam[44,45] or by adopting optimized grating couplers for efficient coupling from free-space to on-chip structures[46]. In addition, we notice that the spontaneous emission process accompanied with the Raman emission offers rich information about electron-phonon interactions that are worth future study, including applications in laser cooling of mechanical resonators[47] and generating entangled photon-phonon pairs. Ultimately, our results represent an important step towards developing chip-integrated quantum circuits and quantum networks that employ multiple solid-state qubits mediated by single photons in a nanophotonic platform.

The authors would like to acknowledge fruitful discussions with Srujan Meesala. This work is supported by Department of Energy (DOE), Laboratory Directed Research and Development program at SLAC National Accelerator Laboratory (contract DE-AC02-76SF00515), Army Research Office (ARO) (W911NF1310309 and W911NF-18-1-0062), Air Force Office of Scientific Research (AFOSR) MURI Center for Quantum Metaphotonics and Metamaterials and MURI for attojoule optoelectronics (Award No. FA9550-17-1-0002), National Science

Foundation (NSF) (ECS-9731293 and DMR-1503759), Stanford Nano Shared Facility, ONR MURI on Quantum Optomechanics (Award No. N00014-15-1-2761), National Science Foundation (NSF) EFRI ACQUIRE program (Award No. 5710004174), and the Army Research Laboratory CDQI (W911NF1520067). Device fabrication is performed in part at the Center for Nanoscale Systems (CNS) at Harvard University, a member of the National Nanotechnology Infrastructure Network (NNIN), which is supported by the National Science Foundation under NSF award No. ECS-0335765.CNS. C.D. acknowledges support from the Andreas Bechtolsheim Stanford Graduate Fellowship.

# Supplementary Materials

# Cavity-enhanced Raman emission from a single color center in a solid

Shuo Sun, Jingyuan Linda Zhang, Kevin A. Fischer, Michael J. Burek,
Constantin Dory, Konstantinos G. Lagoudakis, Yan-Kai Tzeng, Marina Radulaski,
Yousif Kelaita, Amir Safavi-Naeini, Zhi-Xun Shen, Nicholas A. Melosh,
Steven Chu, Marko Lončar, and Jelena Vučković

1. **Derivation of assumptions required for single-excitation regime**

In the discussion for Fig. 1 of the main text, we assume the system contains at most one excitation, so that we can truncate the infinite Jaynes-Cummings ladders to the level structure shown in Fig. 1(b). In the absence of phonon-mediated ground state relaxation, this assumption is always valid. To illustrate why this is the case, we plot the expanded Jaynes-Cummings ladders in Fig. S1(a). We have also denoted all the coherent and incoherent couplings of the system in this figure. Clearly, if the system is initially in the state $|g_1, 0\rangle$, it is not possible to drive the system to the upper ladders formed by $|g_1, 1\rangle$, $|g_2, 2\rangle$ and $|e, 2\rangle$ because there are no couplings that drive a lower-ladder state to the higher ladder. The single-excitation regime is fundamentally guaranteed because we use a laser to drive the transition $|g_1\rangle \leftrightarrow |e\rangle$ which is not coupled to the cavity.

When including the ground state relaxation from $|g_2\rangle$ to $|g_1\rangle$, it is possible to break the single excitation regime. Figure S1(b) illustrates the same expanded energy level structure as shown in Fig. S1(a), but we have included the decay channel from the state $|g_2, n\rangle$ to $|g_1, n\rangle$, where *n* is the number of photons in the cavity. If the ground state relaxation rate $\gamma_{flip}$ is larger than the cavity energy decay rate $\kappa$, then it is possible for the state $|g_2,1\rangle$ to jump to the state $|g_2,2\rangle$ before emitting a photon, by first decaying to the state $|g_1,1\rangle$ with a rate $\gamma_{flip}$ and then rotating to the state $|g_2,2\rangle$ under the coherent drive $\Omega$ and cavity coupling $g$. Therefore when accounting for ground state relaxation, we need the condition $\gamma_{flip} \ll \kappa$ to be satisfied in order to truncate the basis to states with only a single excitation.

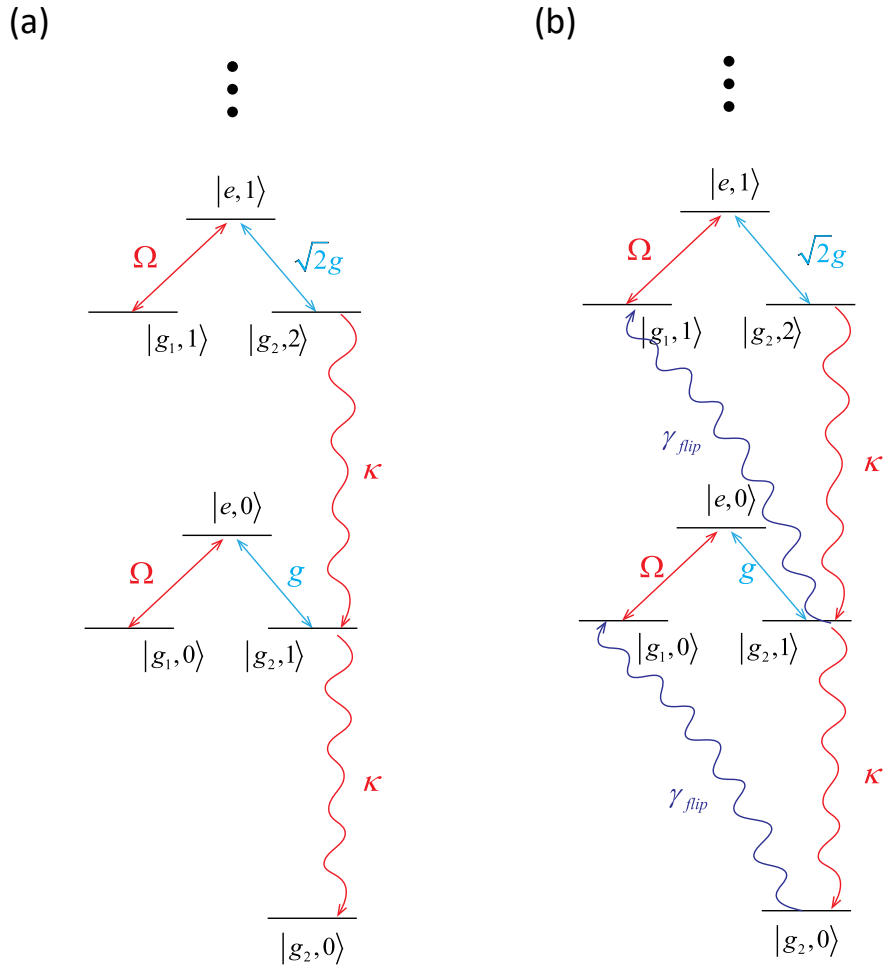

**Figure S1**. System level structure with the expanded Jaynes-Cummings ladders. The panel (a) and (b) show the cases when we ignore and take into account the ground state relaxation respectively.

## 2. Derivation of effective Rabi frequency $\Omega_{eff}$

In this section, we derive the effective Rabi frequency between the state $|g_1,0\rangle$ and the state $|g_2,1\rangle$ after adiabatic elimination of the state $|e,0\rangle$. We focus on the truncated level structure shown in Fig. 1(b) of the main text, and consider only the coherent interactions of the system. These interactions are governed by the Hamiltonian given by

$$\hat{H} = \hbar\Delta|e,0\rangle\langle e,0| + \hbar\frac{\Omega}{2}\left(|e,0\rangle\langle g_1,0| + h.c.\right) + \hbar g\left(|e,0\rangle\langle g_2,1| + h.c.\right). \tag{S1}$$

The equation of motion for the system is governed by the Schrodinger equation, given by $\frac{d}{dt}|\psi(t)\rangle = -\frac{i}{\hbar}\hat{H}|\psi(t)\rangle$, where $|\psi(t)\rangle = c_1(t)|g_1,0\rangle + c_2(t)|g_2,1\rangle + c_3(t)|e,0\rangle$ is the state of the system at time $t$. The coefficients $c_1(t)$, $c_2(t)$, and $c_3(t)$ evolves according to

$$\frac{d}{dt}c_1(t) = -i\frac{\Omega}{2}c_3(t), \tag{S2}$$

$$\frac{d}{dt}c_2(t) = -igc_3(t), \tag{S3}$$

$$\frac{d}{dt}c_3(t) = -i\Delta c_3(t) - i\frac{\Omega}{2}c_1(t) - igc_2(t). \tag{S4}$$

In the limit $\Omega, g \ll \Delta$, the coefficient $c_3(t)$ varies adiabatically with small amplitude such that $|c_3(t)| \ll 1$ and $\frac{d}{dt}c_3(t) \approx 0$ (see Ref. [35] for a rigorous mathematical proof). Then, the evolution of the ground states $|g_1,0\rangle$ and $|g_2,1\rangle$ is independent of the excited state $|e,0\rangle$. This allows solving Eq. S4 for

$$c_3(t) = -\frac{\Omega}{2\Delta}c_1(t) - \frac{g}{\Delta}c_2(t). \tag{S5}$$

Substituting Eq. (S5) back into Eqs. (S2) and (S3), we obtain that

$$\frac{d}{dt}c_1(t) = i\frac{\Omega^2}{4\Delta}c_1(t) + i\frac{\Omega g}{2\Delta}c_2(t), \tag{S6}$$

$$\frac{d}{dt}c_2(t) = i\frac{g^2}{\Delta}c_2(t) + i\frac{\Omega g}{2\Delta}c_1(t). \tag{S7}$$

Therefore, we can treat the system as an effective two-level system comprising the states $|g_1,0\rangle$ and $|g_2,1\rangle$, with effective Hamiltonian

$$\hat{\mathbf{H}}_{eff} = -\hbar\frac{\Omega^2}{4\Delta}|g_1,0\rangle\langle g_1,0| - \hbar\frac{g^2}{\Delta}|g_2,1\rangle\langle g_2,1| - \hbar\frac{\Omega_{eff}}{2}(|g_1,0\rangle\langle g_2,0| + h.c.), \tag{S8}$$

where $\Omega_{eff} = \frac{\Omega g}{\Delta}$ is the effective Rabi frequency.

## 3. Derivation of Raman emission rate

We first calculate the Raman emission rate from a $\Lambda$-system without a cavity. Figure S2(a) shows the energy level structure of the $\Lambda$-system, which consists of two ground states labeled as $|1\rangle$ and $|2\rangle$, and an excited state labeled as $|3\rangle$. We assume that a laser drives the transition $|1\rangle \leftrightarrow |3\rangle$ with a Rabi frequency $\Omega$ and a detuning $\Delta$, and generates Raman emission from the transition $|3\rangle \rightarrow |2\rangle$. In a rotating reference frame with respect to the driving laser, the Hamiltonian of the system is given by

$$\hat{\mathbf{H}} = \hbar\Delta\hat{\boldsymbol{\sigma}}_{33} + \hbar\frac{\Omega}{2}(\hat{\boldsymbol{\sigma}}_{13} + \hat{\boldsymbol{\sigma}}_{31}), \tag{S9}$$

where the operator $\hat{\boldsymbol{\sigma}}_{ij}$ is defined as $\hat{\boldsymbol{\sigma}}_{ij} = |i\rangle\langle j|$ for $i,j \in \{1,2,3\}$. The dynamics of the system expectation values can be derived from the Heisenberg-Langevin equations, resulting in

$$\frac{d\langle\hat{\boldsymbol{\sigma}}_{13}\rangle}{dt} = -(i\Delta + \gamma)\langle\hat{\boldsymbol{\sigma}}_{13}\rangle - i\frac{\Omega}{2}(\langle\hat{\boldsymbol{\sigma}}_{11}\rangle - \langle\hat{\boldsymbol{\sigma}}_{33}\rangle), \tag{S10}$$

$$\frac{d\langle\hat{\boldsymbol{\sigma}}_{33}\rangle}{dt} = -\Gamma_{tot}\langle\hat{\boldsymbol{\sigma}}_{33}\rangle - i\frac{\Omega}{2}(\langle\hat{\boldsymbol{\sigma}}_{31}\rangle - \langle\hat{\boldsymbol{\sigma}}_{13}\rangle), \tag{S11}$$

$$\frac{d\langle\hat{\boldsymbol{\sigma}}_{22}\rangle}{dt} = \Gamma\langle\hat{\boldsymbol{\sigma}}_{33}\rangle. \tag{S12}$$

In Eqs. (S10) - (S12), $\gamma$ is the dipole decoherence rate of transition $|1\rangle \leftrightarrow |3\rangle$, $\Gamma$ is the spontaneous emission rate of the transition $|3\rangle \rightarrow |2\rangle$, $\Gamma_{tot}$ is the total decay rate of the excited state $|3\rangle$. In the large detuning limit where $\Omega, \Gamma_{tot}, \gamma \ll \Delta$, the excited state $|3\rangle$ is weakly excited. Thus, we can adiabatically eliminate this state by taking the steady-state solution of Eqs. (S10) and (S11). Substituting the steady-state solutions into Eq. (S12), we obtain that

$$\frac{d\langle\hat{\boldsymbol{\sigma}}_{22}\rangle}{dt} = \Gamma \cdot \frac{\alpha}{2\alpha + \Gamma_{tot}}\left(1 - \langle\hat{\boldsymbol{\sigma}}_{22}\rangle\right), \tag{S13}$$

where $\alpha = \frac{\Omega^2}{2} \cdot \frac{\gamma}{\Delta^2 + \gamma^2}$. To obtain Eq. (S13), we used the identity that $\langle\hat{\boldsymbol{\sigma}}_{11}\rangle + \langle\hat{\boldsymbol{\sigma}}_{22}\rangle + \langle\hat{\boldsymbol{\sigma}}_{33}\rangle = 1$ and $\langle\hat{\boldsymbol{\sigma}}_{31}\rangle = \langle\hat{\boldsymbol{\sigma}}_{13}\rangle^\dagger$.

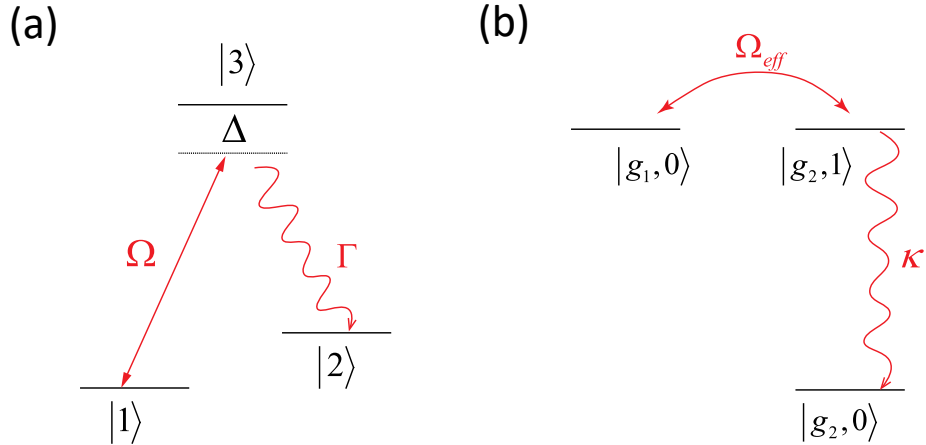

**Figure S2**. Schematics of Raman emission process for a bare emitter (a) and an emitter that couples to the cavity (b) respectively.

Eq. (S13) has a very clear physical interpretation. It shows that the population of the ground state $|2\rangle$ grows exponentially with a rate $R_0 = \Gamma \cdot \dfrac{\alpha}{2\alpha + \Gamma_{tot}}$, which is exactly the Raman emission rate. To simplify the expression of the Raman emission rate, we assume the ideal scenario where the excited state $|3\rangle$ decays only to the ground state $|2\rangle$ through spontaneous emission ($\Gamma_{tot} = \Gamma$), and the linewidth of transition $|1\rangle \leftrightarrow |3\rangle$ is lifetime limited ($\gamma = \dfrac{\Gamma}{2}$). This assumption gives the upper bound of the Raman emission rate. Under this assumption, the Raman emission rate is given by $R_0 = \Gamma \cdot \dfrac{\alpha}{2\alpha + \Gamma}$, where $\alpha = \dfrac{\Omega^2}{4} \cdot \dfrac{\Gamma}{\Delta^2 + (\Gamma/2)^2}$.

In the large detuning limit where $\Omega, \Gamma \ll \Delta$, we have $\alpha \ll \Gamma$, thus the Raman emission rate is given by $R_0 = \alpha = \dfrac{\Omega^2}{4\Delta^2}\Gamma$.

Now we calculate the rate of the cavity enhanced Raman emission. As explained in the main text, by adiabatic elimination of the state $|e,0\rangle$, the cavity enhanced Raman emission can be understood as an effective Rabi oscillation between $|g_1,0\rangle$ and $|g_2,1\rangle$, followed by an decay from $|g_2,1\rangle$ to $|g_2,0\rangle$ with a rate $\kappa$. Figure S2(b) shows this simplified picture. This picture resembles the three-level systems shown in Fig. S2(a), if we define $|1\rangle \equiv |g_1,0\rangle$, $|2\rangle \equiv |g_2,0\rangle$ and $|3\rangle \equiv |g_2,1\rangle$. Therefore, we could calculate the cavity enhanced Raman emission rate $R_c$ following the same derivations shown above, given by $R_c = \dfrac{\Omega_{eff}^2}{\kappa}$, where $\Omega_{eff}$ is given by $\Omega_{eff} = \Omega g/\Delta$ as shown in Sec. 2.

## 4. Details of measurement setups and techniques

### 4.1 Complete schematic of the measurement setup

Figure S3 shows a schematic of the optical setup we used for all our reported measurements. We mount the sample in a closed-cycle cryostat (Montana instruments) and cool it down to 4 K. We use a confocal microscope with an objective lens that has a numerical aperture of 0.9 to perform sample excitation and collection. We use three light sources for sample excitation: a supercontinuum source which is used for characterizing the transmission spectrum of the cavity (Fig. 2(b) of the main text), a pulsed Ti: Sapphire laser which is used for lifetime measurement (Fig. 2(d) of the main text), and a tunable continuous-wave Ti: Sapphire laser which is used both for generating photoluminescence (Fig. 2(c) of the main text) and Raman emission (Fig. 3 of the main text). We tune the wavelength of the continuous-wave Ti: Sapphire laser to 720 nm to generate photoluminescence, and tune it to near transition $|g_1\rangle \leftrightarrow |e\rangle$ to generate Raman emission. We couple the collected signal either to a single mode fiber, or directly to a spectrometer through free-space. We use a monochromator to reject the laser reflection for the Raman emission measurements.

### 4.2 Cavity transmission measurement

To obtain the cavity transmission spectrum as shown in Fig. 2(b) of the main text, we use a broadband supercontinuum laser to excite the notch located at one end of the nanobeam. The notch is designed to couple light from free-space into the waveguide (see Sec. 8 for detailed information). The second notch at the other end of the nanobeam scatters the transmitted signal back into free space, which we collect with a single mode fiber. We send the fiber-coupled signal to the spectrometer to record the cavity transmission spectrum.

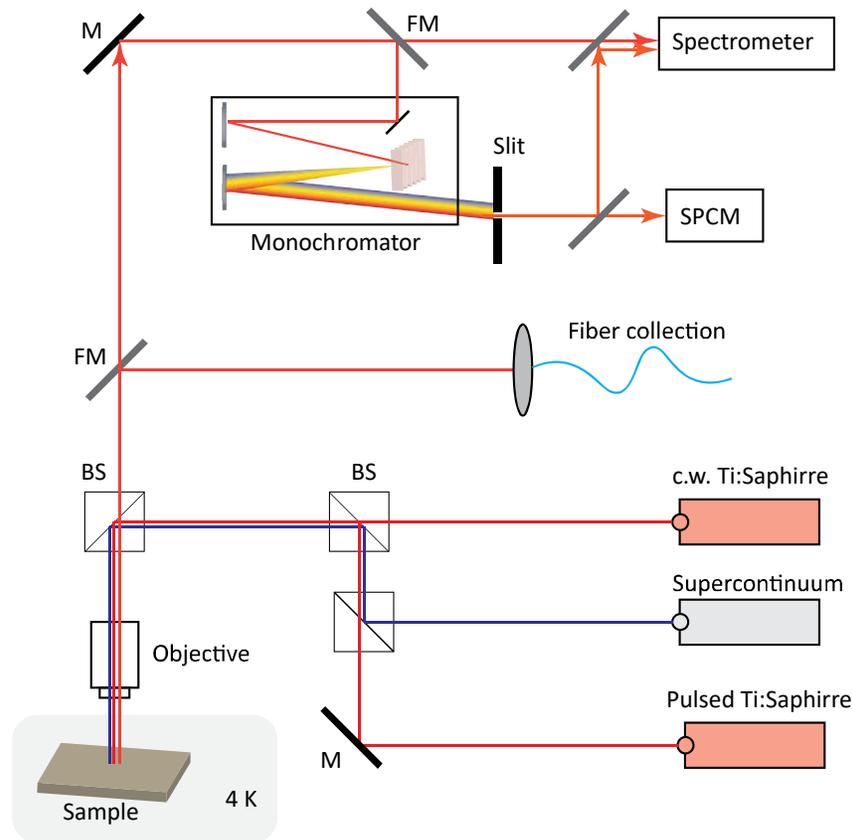

**Figure S3**. Complete schematics of the measurement setup. M, mirror; FM, flip mirror; SPCM, single-photon counting module.

**4.3 Photoluminescence spectrum measurement**

Figure 2(c) of the main text shows the measured photoluminescence spectrum of the $SiV^-$ center. We directly excite the $SiV^-$ center at the center of the nanobeam using a Ti: Sapphire continuous-wave laser that is tuned to 720 nm, which is blue detuned from all the four optical transitions of the SiV¯ center (ranging from 735 to 738 nm). This is a typical way to generate photoluminescence from SiV¯ centers [13, 31, 36]. To eliminate the effect of the cavity on the emission properties, we red-detune the cavity by more than 40 cavity linewidths from all the four transitions of the SiV¯ center. We couple the collected signal directly to a spectrometer through free-space. We use a dielectric-coated band pass filter with a center wavelength of 740 nm and a bandwidth of 13 nm to spectrally reject the laser reflection before the spectrometer.

### 4.4 Time-resolved photoluminescence measurement

Figure 2(d) of the main text shows the time-resolved photoluminescence measurement which we use to extract the lifetime of the excited state. In this measurement, we use a 2-ps pulse with a center wavelength of 720 nm to excite the SiV¯ center at the center of the nanobeam, and detect the photoluminescence emission with a silicon avalanche photodiode single-photon detector. We use a monochromator to spectrally filter only the emission from the transition $|g_2\rangle \leftrightarrow |e\rangle$ before the detector. The data shown in Fig. 2(d) of the main text is generated by a time-correlated single-photon counting system (PicoHarp 300) which records the delay time between the laser pulse and each detected photon.

### 4.5 Cavity-enhanced Raman emission measurement

In the cavity-enhanced Raman emission measurement (Fig. 3 of the main text), we excite the device using a Ti: Sapphire continuous-wave laser that is near resonant with transition $|g_1\rangle \leftrightarrow |e\rangle$, and collect the emission from the cavity. We couple the excitation laser through the notch of the waveguide, which is designed to couple light from free-space into the waveguide (see Sec. 8 for detailed information). We collect the cavity emission from its scattering into free-space. We couple the signal into a monochromator to spectrally reject the excitation laser, and then send it to a spectrometer (which is essentially another monochromator followed by a CCD camera).

As depicted in Fig. S3, the mechanism of the monochromator is based on spatial dispersion of light upon diffraction from a grating. After spatial dispersion, a mechanical slit is used to spectrally select the signal within certain spectral range. The filter bandwidth and center wavelength can be tuned by adjusting the width and the position of the slit. In our measurement, we align the monochromator filter to be centered at the transition $|g_2\rangle \leftrightarrow |e\rangle$, with a filter bandwidth of 120 GHz. We measure a maximum transmission efficiency of the monochromator to be 17.4%. The transmission efficiency drops to 0.0014% at the frequency of the laser, which is ~544 GHz blue detuned from the center of the spectral filter. Therefore the extinction ratio of the monochromator is determined to be 12400.

## 5. Photoluminescence of bulk SiV⁻ center ensembles

Figure S4 shows the photoluminescence spectrum of an ensemble of SiV⁻ centers in the bulk diamond. Here we plot the emission spectrum as a function of detuning from the peak C.

From the frequency splitting between the emission peaks C and D, we calculate the ground state splitting to be $\delta_g/2\pi = 50\,\text{GHz}$ for bulk SiV⁻ centers. This value is consistent with the values reported in many studies [13, 30-32, 36].

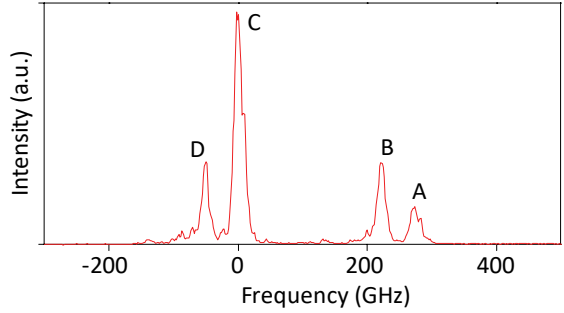

**Figure S4**. Photoluminescence spectrum of an ensemble of SiV⁻ centers in the bulk diamond.

## 6. Second order correlation measurements

We perform second order correlation measurements to verify that the emission peaks C and D shown in Fig. 2(c) of the main text originate from a single silicon-vacancy (SiV⁻) center. We excite the device using a 2-ps pulsed laser with a repetition rate of 80 MHz and a center wavelength of 720 nm. We collect the emission using a multi-mode fiber and send it to a Hanbury Brown-Twiss (HBT) intensity interferometer composed of a 50/50 beam-splitter and two Single Photon Counting Modules (SPCMs). We use a time correlated single-photon counting system (PicoHarp 300) to process the detection events from the two SPCMs and obtain the second order correlation. To isolate the emission from peaks C and D respectively, we resonantly couple the cavity with either peak C or D and use the cavity as a spectral filter. Figure S5 shows the measured second order correlations when the cavity is resonant with peak C and D respectively, where $\tau$ is the delay time between two detection events obtained by the two SPCMs. We observe strong suppression of the second order correlation near $\tau = 0$ in both cases, confirming that both emission peaks are from a single SiV¯ center.

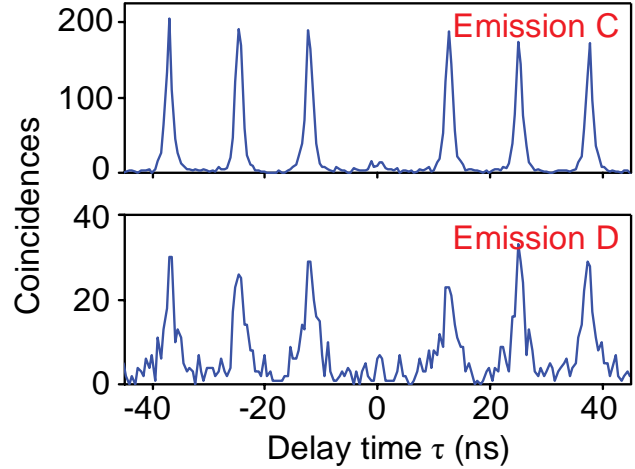

**Figure S5**. Second order correlations of the SiV¯ emission when the cavity is resonant with emission peak C and D respectively.

## 7. Estimation of Purcell factor

We estimate the Purcell factor $F$ based on the calculated coupling strength, defined as $F = \dfrac{4g^2/\kappa}{\Gamma_{bare}}$, where $\Gamma_{bare}$ is the spontaneous emission rate of transition $|e\rangle \to |g_2\rangle$ when it decouples with the cavity. It is not straightforward to obtain the value of $\Gamma_{bare}$, since we can only measure the lifetime (or total decay rate) of the excited state $|e\rangle$ when the cavity is far detuned, which includes decay through non-radiative processes, phonon-sideband emissions, and spontaneous emission into the other zero-phonon line $|e\rangle \to |g_1\rangle$. To obtain a lower bound of the

Purcell factor, here we estimate a higher bound of $\Gamma_{bare}$. We assume a higher-bound quantum yield of $\eta_{radiative} = 30\%$ and a zero-phonon-line emission fraction of $\eta_{ZPL} = 80\%$. These two values together have accounted for the fraction of decay through emissions into the zero-phonon-lines. However, the excited state $|e\rangle$ could still decay through two possible ZPLs, $|e\rangle \to |g_1\rangle$ and $|e\rangle \to |g_2\rangle$. We estimate the fraction of the zero-phonon-line emission into transition $|e\rangle \to |g_2\rangle$ to be $\eta_D = 10\%$ based on the measured photoluminescence spectrum shown in Fig. 2(c) of the main text. Therefore, we could calculate the higher-bound of $\Gamma_{bare}$ given by $\Gamma_{bare} = \eta_{radiative} \cdot \eta_{ZPL} \cdot \eta_D \frac{1}{\tau_{off}} = 2\pi \times 2.1\,\text{MHz}$. We thus estimate the lower-bound of the Purcell factor to be $F = \frac{4g^2/\kappa}{\Gamma_{bare}} = 22.7$.

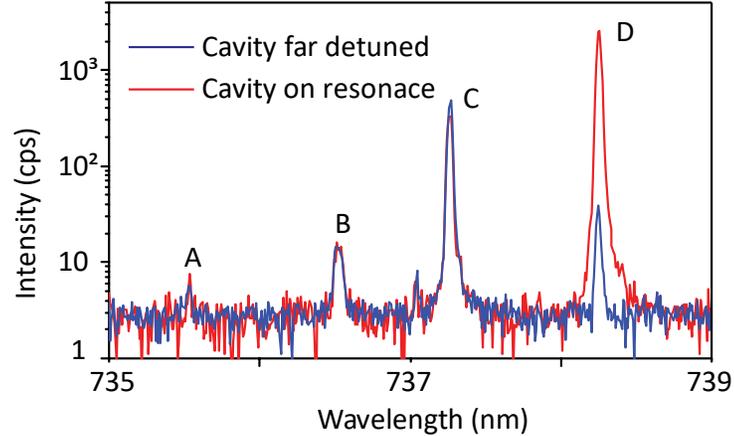

**Figure S6**. Photoluminescence spectrum of the emitter when the cavity is resonant (red) and far detuned (blue) from the transition $|e\rangle \to |g_2\rangle$.

To further confirm the lower-bound value of the Purcell factor, we measure the photoluminescence of the emitter when the cavity is on resonance and far detuned from transition $|e\rangle \to |g_2\rangle$, shown as the red and blue solid lines in Fig. S6. The emission intensity from transition $|e\rangle \to |g_2\rangle$ (peak D) increases by a factor of 60 when the cavity is resonant, which is another supportive evidence of a strong Purcell enhancement.

## 8. Design of the notches for free-space to waveguide coupling

In our experiment, to reject the direct reflection of the laser from the sample surface, we spatially separate the excitation and collection by irradiating the laser at a notch located at the end of the nanobeam, which is designed for free-space to waveguide coupling. Figure S7(a) shows the dimensions of the notch, which follows a similar design as shown in Ref. [13]. Figure S7(b) shows a scanning electron microscope image of the notch after fabrication.

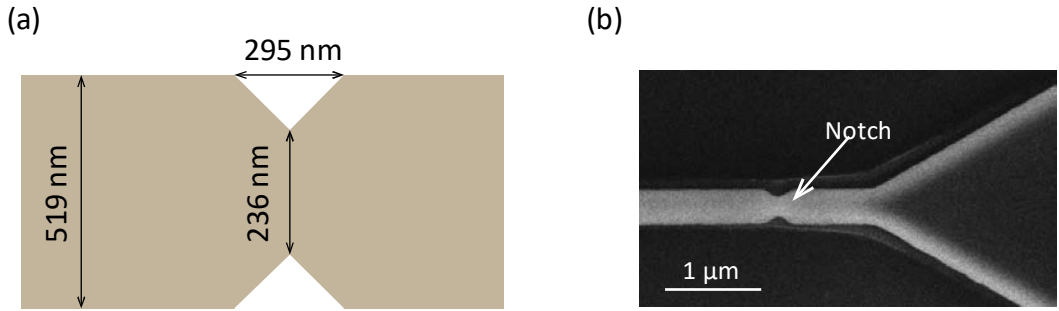

**Figure S7**. (a) Dimensions of the notch that is designed for coupling light from free-space into the nanobeam waveguide. (b) Scanning electron microscope image of the notch after fabrication.

## 9. Complete derivation of system master equation

We first derive the Lindblad operator for the electron-phonon interaction Hamiltonian $\hat{H}_{sys-bath}$. To do that we have to first write the interaction Hamiltonian $\hat{H}_{sys-bath}$ in the diagonal basis of $\hat{\mathbf{H}}_{sys}$ (Eq. (1) of the main text), and then write it in the rotating reference frame with respect to $\hat{\mathbf{H}}_{sys} + \hat{\mathbf{H}}_{bath}$. In the diagonal basis, we can write $\hat{\mathbf{H}}_{sys}$ as

$$\hat{\mathbf{H}}_{sys} = \omega_+ |+\rangle\langle+| + \omega_- |-\rangle\langle-| + \omega_d |d\rangle\langle d|, \tag{S14}$$

where the eigenstates $|+\rangle$, $|-\rangle$, and $|d\rangle$ are given by Eqs. (5) – (7) in the main text, the eigenfrequencies $\omega_+$, $\omega_-$, and $\omega_d$ are given by $\omega_+ = \Delta + \frac{(\Omega/2)^2 + g^2}{\Delta}$, $\omega_- = -\frac{(\Omega/2)^2 + g^2}{\Delta}$, and $\omega_d = 0$ respectively. In the rotating reference frame with respect to $\hat{\mathbf{H}}_{sys} + \hat{\mathbf{H}}_{bath}$, we could rewrite $\hat{\mathbf{H}}_{sys-bath}$ as

$$\hat{\mathbf{H}}_{sys-bath} = \frac{\sqrt{g^2 + (\Omega/2)^2}}{\Delta} \sum_{\mathbf{k}} \mathbf{b}_{\mathbf{k}} \left( x_{\mathbf{k}} |+\rangle\langle-| e^{i(\Lambda_1 - \omega_{\mathbf{k}})t} + y_{\mathbf{k}} |+\rangle\langle d| e^{i(\Lambda_2 - \omega_{\mathbf{k}})t} \right) + h.c., \tag{S15}$$

where $\Lambda_1$ and $\Lambda_2$ are given by $\Lambda_1 = \omega_+ - \omega_-$ and $\Lambda_2 = \omega_+ - \omega_d$ respectively, $x_{\mathbf{k}}$ and $y_{\mathbf{k}}$ are given by $x_{\mathbf{k}} = p_{\mathbf{k}} \frac{(\Omega/2)^2}{g^2 + (\Omega/2)^2} + q_{\mathbf{k}} \frac{g^2}{g^2 + (\Omega/2)^2} - r_{\mathbf{k}}$ and $y_{\mathbf{k}} = (p_{\mathbf{k}} - q_{\mathbf{k}}) \frac{g(\Omega/2)}{g^2 + (\Omega/2)^2}$ respectively. To obtain Eq. (S15), we have utilized the rotating wave approximation to keep only the slowly varying terms. We eliminate the phonon coupling terms with the operators $|+\rangle\langle+|$, $|-\rangle\langle-|$, and $|d\rangle\langle d|$ because they interact with phonons at zero frequency where phonon density of states vanishes. Similarly, we eliminate the phonon coupling terms with the operators $|d\rangle\langle-|$, and $|-\rangle\langle d|$ because they interact with phonons at a low frequency near $\omega_d - \omega_- = \frac{(\Omega/2)^2 + g^2}{\Delta}$, which is in the order of 100 MHz. Such frequencies correspond to a phonon wavelength longer than 10 μm, which cannot exist in our nanobeam structure.

Now we can derive a master equation by integrating the von Neumann equation for the

density matrix $\rho$ of the joint system and phonon bath, and then tracing over the phonon modes, given by

$$\frac{d\rho_{sys}}{dt} = -\int_0^t \text{tr}_{bath}\left(\left[\hat{\mathbf{H}}_{sys-bath}(t),\left[\hat{\mathbf{H}}_{sys-bath}(t'),\rho(t')\right]\right]\right)dt'. \tag{S16}$$

We make the Born-Markov approximation, which allows us to substitute $\rho(t')$ with $\rho(t)$ and write it as $\rho = \rho_{sys} \otimes \rho_{bath}$. These assumptions result in a master equation given by

$$\frac{d\rho_{sys}}{dt} = -\int_0^t \text{tr}_{bath}\left(\left[\hat{\mathbf{H}}_{sys-bath}(t),\left[\hat{\mathbf{H}}_{sys-bath}(t),\rho_{sys}(t)\otimes\rho_{bath}\right]\right]\right)dt'. \tag{S17}$$

We further rewrite $\hat{\mathbf{H}}_{sys-bath}$ as $\hat{\mathbf{H}}_{sys-bath} = \hat{\mathbf{H}}^{(1)}_{sys-bath} + \hat{\mathbf{H}}^{(2)}_{sys-bath}$, where $\hat{\mathbf{H}}^{(1)}_{sys-bath}$ and $\hat{\mathbf{H}}^{(2)}_{sys-bath}$ are given by

$$\hat{\mathbf{H}}^{(1)}_{sys-bath} = \frac{\sqrt{g^2+(\Omega/2)^2}}{\Delta}\sum_{\mathbf{k}} x_{\mathbf{k}}\mathbf{b}_{\mathbf{k}}|+\rangle\langle-|e^{i(\Lambda_1-\omega_{\mathbf{k}})t} + h.c., \tag{S18}$$

$$\hat{\mathbf{H}}^{(2)}_{sys-bath} = \frac{\sqrt{g^2+(\Omega/2)^2}}{\Delta}\sum_{\mathbf{k}} y_{\mathbf{k}}\mathbf{b}_{\mathbf{k}}|+\rangle\langle d|e^{i(\Lambda_2-\omega_{\mathbf{k}})t} + h.c.. \tag{S19}$$

Since $\hat{\mathbf{H}}^{(1)}_{sys-bath}$ and $\hat{\mathbf{H}}^{(2)}_{sys-bath}$ involve interaction with phonons of different frequencies separated by ~100 MHz, they cannot interact with the same phonon mode. Therefore, we could further rewrite Eq. (S17) as

$$\frac{d\rho_{sys}}{dt} = -\sum_{m=1}^{2}\int_0^t \text{tr}_{bath}\left(\left[\hat{\mathbf{H}}^{(m)}_{sys-bath},\left[\hat{\mathbf{H}}^{(m)}_{sys-bath},\rho_{sys}(t)\otimes\rho_{bath}\right]\right]\right)dt'. \tag{S20}$$

This leads to the final master equation given by $\frac{d\rho_{sys}}{dt} = -i\left[\hat{\mathbf{H}}_{sys},\rho_{sys}\right] + \sum_{m=1}^{2} L^{(m)}_{phonon}(\rho_{sys})$, where $L^{(1)}_{phonon}(\rho_{sys})$ and $L^{(2)}_{phonon}(\rho_{sys})$ are given by

$$L^{(1,2)}_{phonon}(\rho_{sys}) = \frac{g^2 + (\Omega/2)^2}{\Delta^2} J_{1,2}(\Lambda_{1,2}) \left[ n_{th}(\Lambda_{1,2}) D(|+\rangle\langle-|) + (1 + n_{th}(\Lambda_{1,2})) D(|-\rangle\langle+|) \right], \quad (S21)$$

where $J_{1,2}(\Lambda_{1,2})$ is the phonon spectral density given by $J_1(\Lambda_1) = 2\pi \sum_{\mathbf{k}} |x_{\mathbf{k}}|^2 \delta(\omega_{\mathbf{k}} - \Lambda_1)$ and $J_2(\Lambda_2) = 2\pi \sum_{\mathbf{k}} |y_{\mathbf{k}}|^2 \delta(\omega_{\mathbf{k}} - \Lambda_2)$ respectively, and $n_{th}(\Lambda_{1,2})$ is the number of phonons per mode, which follows the Bose-Einstein distribution given by $n_{th}(\Lambda_{1,2}) = [\exp(\Lambda_{1,2}/k_B T) - 1]^{-1}$. Since both the phonon spectral density $J_{1,2}(\Lambda_{1,2})$ and the thermal distribution function $n_{th}(\Lambda_{1,2})$ are relatively flat as a function of phonon frequency, and $\Lambda_{1,2} \approx \Delta$ in the limit $\Omega, g \ll \Delta$, we could approximately write $J_{1,2}(\Lambda_{1,2})$ and $n_{th}(\Lambda_{1,2})$ as $J_{1,2}(\Lambda_{1,2}) = J_{1,2}(\Delta)$ and $n_{th}(\Lambda_{1,2}) = n_{th}(\Delta)$ respectively. This gives the phonon dissipator provided as Eq. (4) in the main text.

In our numerical simulation, we use the full master equation that account for all possible dissipation mechanisms, given by $\frac{d\rho_{sys}}{dt} = -i[\hat{\mathbf{H}}_{sys}, \rho_{sys}] + L_{phonon}(\rho_{sys}) + L_{cav}(\rho_{sys}) + L_{SiV}(\rho_{sys})$, where $L_{cav}(\rho_{sys}) = \kappa D(\hat{\mathbf{a}})$ is the cavity decay, and $L_{SiV}(\rho_{sys})$ is the decay of the SiV$^-$ center, given by

$$L_{SiV}(\rho_{sys}) = \gamma_1 D(|g_1\rangle\langle e|) + \gamma_2 D(|g_2\rangle\langle e|) + \gamma_{flip} D(|g_1\rangle\langle g_2|). \quad (S22)$$

In Eq. (S22), $\gamma_1$ and $\gamma_2$ are decay rates from the excited state $|e\rangle$ to the ground states $|g_1\rangle$ and $|g_2\rangle$ respectively, $\gamma_{flip}$ is the decay rate from $|g_2\rangle$ to $|g_1\rangle$. We do not include the state flipping from $|g_1\rangle$ to $|g_2\rangle$ in the Liouvillian superoperator, because this process requires absorption of a phonon at $\delta_g/2\pi = 544 \text{ GHz}$ that is much larger than the value $k_B T/2\pi = 83 \text{ GHz}$, and therefore is much slower than its reverse process.

We numerically solve the master equation of the system and calculate the cavity emission spectrum using the quantum regression theorem. We fix $g$ and $\kappa$ using experimentally measured values given by $g/2\pi = 0.80\,\text{GHz}$ and $\kappa/2\pi = 53.7\,\text{GHz}$. We assume that the decay rates from the excited state $|e\rangle$ to the ground states $|g_1\rangle$ and $|g_2\rangle$ are identical, thus we have $\gamma_1 = \gamma_2 = \dfrac{1}{2\tau_{off}} = 2\pi \times 0.046\,\text{GHz}$. This assumption is valid since the dominant decay mechanism of the excited state is through a non-radiative process [37], which rate is irrelevant of the final ground state. To determine the driving Rabi frequency $\Omega$, we resonantly drive transition $|g_1\rangle \leftrightarrow |e\rangle$, and measure the fluorescence intensity as we vary the driving laser power. This measurement allows us to obtain the saturation power for transition $|g_1\rangle \leftrightarrow |e\rangle$, enabling us to determine the driving Rabi frequency based on the measured laser power. In our experiment, we use a driving Rabi frequency of $\Omega/2\pi = 2.58\,\text{GHz}$. The only parameter we cannot determine is $\gamma_{flip}$. However, as we verified numerically, the value of $\gamma_{flip}$ only determines the number of excitation and emission cycles per second – it does not affect the R/S ratio. In the calculation we simply fix $\gamma_{flip}$ to be $\gamma_{flip}/2\pi = 0.8\,\text{GHz}$ based on an estimate from a previous literature [31].